\documentclass[acmtog]{acmart}

\usepackage{booktabs} 

\usepackage{subcaption} 

\usepackage[ruled]{algorithm2e} 

\SetAlFnt{\small}
\SetAlCapFnt{\small}
\SetAlCapNameFnt{\small}
\SetAlCapHSkip{0pt}

\begin{document}
\title{DreamPrinting: Volumetric Printing Primitives for High-Fidelity 3D Printing}


\author{Youjia Wang}
\affiliation{%
 \institution{ShanghaiTech University}
 \country{China}}
 \affiliation{%
 \institution{LumiAni Technology}
 \country{China}}
\email{wangyj2@shanghaitech.edu.cn}
\authornote{Equal contributions.}

\author{Ruixiang Cao}
\affiliation{%
 \institution{Kyoto University}
 \country{Japan}}
\email{cao.ruixiang.65h@st.kyoto-u.ac.jp}
\authornotemark[1]
\authornote{Concept originator.}

\author{Teng Xu}
\affiliation{%
 \institution{ShanghaiTech University}
 \country{China}}
 \affiliation{%
 \institution{LumiAni Technology}
 \country{China}}
\email{xt@shanghaitech.edu.cn}

\author{Yifei Liu}
\affiliation{%
 \institution{ShanghaiTech University}
 \country{China}}
 \affiliation{%
 \institution{LumiAni Technology}
 \country{China}}
\email{arnoliu@shanghaitech.edu.cn}

\author{Dong Zhang}
\affiliation{%
 \institution{ShanghaiTech University}
 \country{China}}
 \affiliation{%
 \institution{LumiAni Technology}
 \country{China}}
\email{zhangdong@shanghaitech.edu.cn}

\author{Yiwen Wu}
\affiliation{%
 \institution{ShanghaiTech University}
 \country{China}}
 \affiliation{%
 \institution{LumiAni Technology}
 \country{China}}
\email{wuyw2023@shanghaitech.edu.cn}

\author{Jingyi Yu}
\affiliation{%
 \institution{ShanghaiTech University}
 \country{China}}
\email{yujingyi@shanghaitech.edu.cn}
\authornote{Corresponding author.}



\begin{teaserfigure} 
    \centering\includegraphics[width=0.9\textwidth]{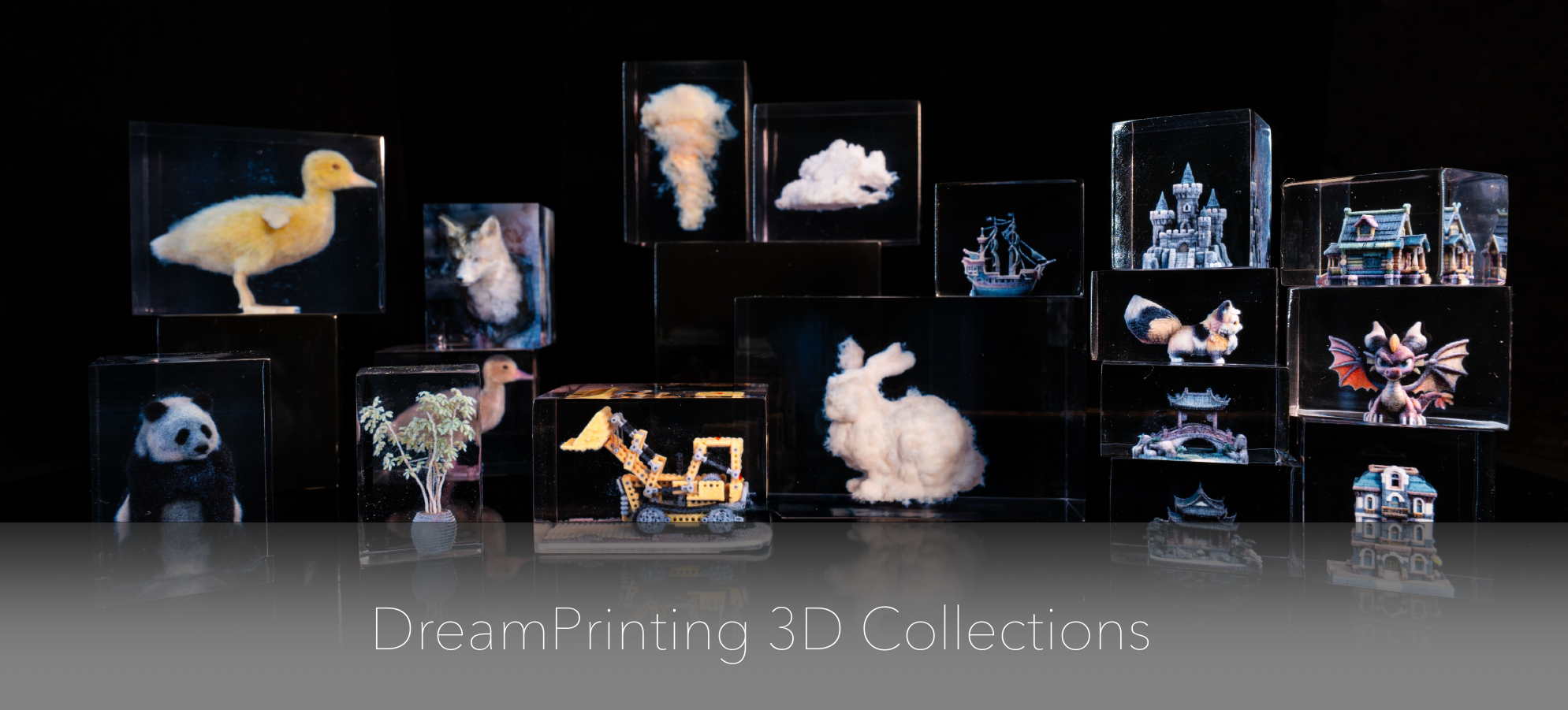} \caption{
    We present \textit{DreamPrinting}, a novel 3D volumetric printing pipeline that precisely assigns pigment labels at the voxel level, enabling the reproduction of complex visual effects with remarkable realism. By capturing both geometric detail and vibrant color fidelity, our method transforms radiance-based 3D assets into tangible prints that faithfully reflect real-world appearances, bridging the gap between the virtual and the physical with unprecedented fidelity.
} 
    \label{fig:teaser} 
\end{teaserfigure}

\begin{abstract}
Translating the rich visual fidelity of volumetric rendering techniques into physically realizable 3D prints remains an open challenge. We introduce DreamPrinting, a novel pipeline that transforms radiance-based volumetric representations into explicit, material-centric Volumetric Printing Primitives (VPPs). While volumetric rendering primitives (e.g., NeRF) excel at capturing intricate geometry and appearance, they lack the physical constraints necessary for real-world fabrication, such as pigment compatibility and material density. DreamPrinting addresses these challenges by integrating the Kubelka-Munk model with a spectrophotometric calibration process to characterize and mix pigments for accurate reproduction of color and translucency. The result is a continuous-to-discrete mapping that determines optimal pigment concentrations for each voxel, ensuring fidelity to both geometry and optical properties. A 3D stochastic halftoning procedure then converts these concentrations into printable labels, enabling fine-grained control over opacity, texture, and color gradients. Our evaluations show that DreamPrinting achieves exceptional detail in reproducing semi-transparent structures—such as fur, leaves, and clouds—while outperforming traditional surface-based methods in managing translucency and internal consistency. Furthermore, by seamlessly integrating VPPs with cutting-edge 3D generation techniques, DreamPrinting expands the potential for complex, high-quality volumetric prints, providing a robust framework for printing objects that closely mirror their digital origins.

\end{abstract}

%
%

%
%


\maketitle

\section{Introduction}
Three-dimensional (3D) printing has emerged as a transformative technology that empowers individuals and industries alike to materialize highly specific and unique concepts across diverse fields such as art, fashion, architecture, and engineering. By facilitating unparalleled customization, it provides a vehicle to bring everyone’s dreams to life, whether grand or small, opaque or translucent, transforming the imagined into the tangible with exquisite precision.

Existing 3D printing methodologies, such as Fused Deposition Modeling (FDM), predominantly rely on surface-based representations like meshes to delineate object boundaries efficiently. FDM builds objects layer by layer by extruding thermoplastic materials through a heated nozzle, allowing for the creation of intricate geometries and internal cavities. However, despite its widespread adoption and mechanical stability, FDM faces significant constraints in capturing fine details and accurately resolving opacity due to physical limitations of deposition processes and nozzle sizes, which restrict the minimum feature size and hinder the reproduction of highly intricate structures.

Advancements in full-color inkjet 3D printing aim to overcome the limitations of traditional methods by enabling ultra-fine geometric details and vibrant color reproduction. These systems convert mesh-based models into surface voxels (texels), allowing precise control over color and material distribution. Unlike traditional 3D printing, which typically uses a fixed set of materials and relies on post-processing for color application, inkjet printing employs a broader range of pigments to directly achieve intricate color gradients and translucency, simulating real-world materials like glass or fabric. However, few approaches have effectively utilized transparency and complex pigment combinations to reproduce fine details, such as delicate geometries and textures like fur. The core challenge lies in the lack of suitable primitives that integrate volumetric printing with the physical requirements of 3D printing, such as material density, opacity, and light transmission.

\begin{figure*}
  \includegraphics[width=\linewidth]{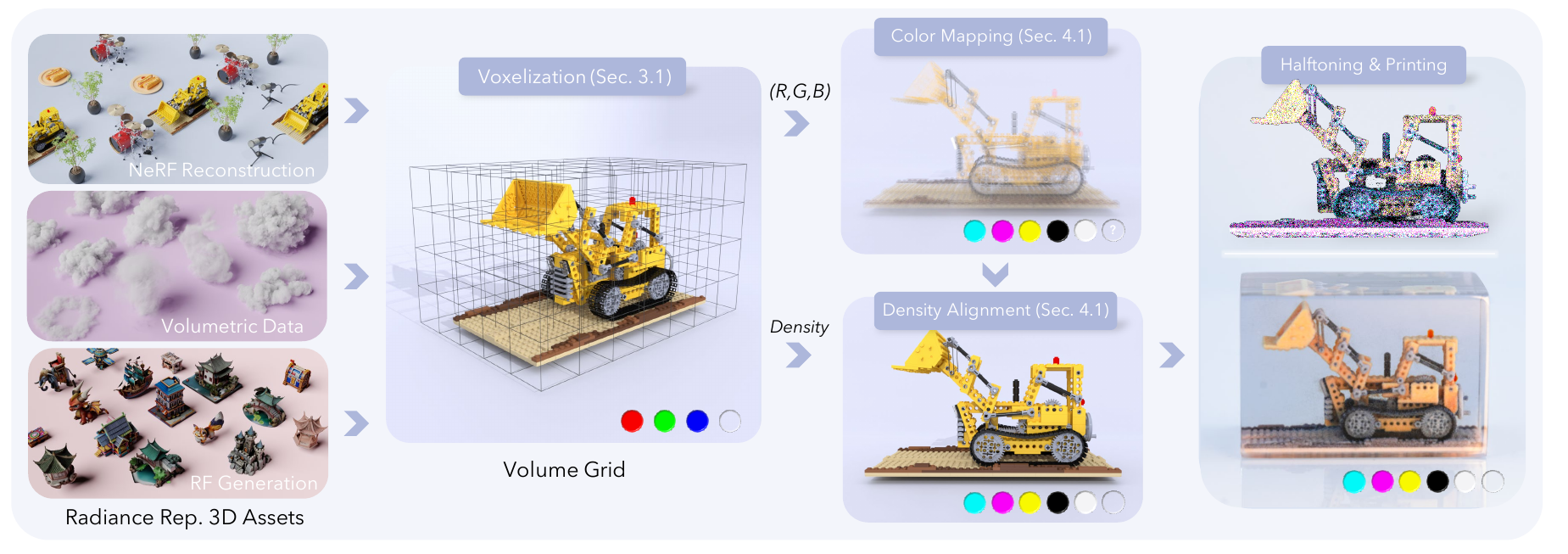}
  \caption{Overview of our pipeline for transforming radiance representation into material-centric representation using Volumetric Printing Primitives (VPPs). VPPs are positioned on a uniform grid, storing RGB color and density values to form the pre-print Volume. Color Mapping converts RGB to pigment (\texttt{C, M, Y, K, W}) concentrations, while Density Alignment modifies \texttt{Cl} or \texttt{K}/\texttt{W} concentrations to match target densities. A Stochastic Halftoning Module then assigns discrete pigment labels for direct 3D printing.}
  \label{fig:pipeline}
\end{figure*}

In 3D modeling and rendering, recent volumetric rendering primitives (VRPs) such as Neural Radiance Fields (NeRF) \cite{mildenhall2021nerf}  and 3D Gaussian Splatting (3DGS)\cite{Kerbl20233DGS} have emerged as an alternative to 3D meshes for high quality modeling and rendering. By employing an additional density channel, they not only enable efficient capture and rendering of intricate geometry and textures without extracting polygonal meshes but also offer smooth workflows for dynamic and immersive applications like gaming, VR/AR, and visual effects, pushing the boundaries of what’s possible in real-time and pre-rendered graphics. 

In this paper, we introduce Volumetric Printing Primitives (VPPs) to bridge the gap between volumetric rendering and volumetric printing. First, we observe that the printing primitives differ fundamentally from the rendering ones in how they encode and manipulate data. While VRPs can freely combine RGB and density values, VPPs are constrained by the need to assign each voxel a single pigment. Then, the mechanical design of 3D printer nozzles imposes further limitations. Even with the most advanced ones, by far only a limited number of pigments can be used in a single printing task, each corresponding to fixed density and material properties. Furthermore, the physical attributes of pigments—such as their reflection and transmission rates—are unknown and require a calibration process to characterize their behavior. Finally and most importantly, the ray integration process used in VRPs, where RGB and density are aggregated along light paths, is unsuitable for real-world pigments because pigments exhibit different transmission rates across wavelengths of light. Together, these challenges render the computational logic of VRPs incompatible with the physical constraints of VPPs.

Inspired by~\cite{abdul2005spectral,sochorova2021practical, cao2025poxelvoxelreconstruction3d}, we present a novel 3D volumetric printing pipeline called DreamPrinting that resolves each mismatch between VRPs and VPPs. In a nutshell, DreamPrinting aims to assign pigment labels to every voxel in the VPP, enabling it to achieve effects approximating real-world visual fidelity. Theoretically, this is a combinatorial labeling problem: finding the optimal pigment combination such that the integrated printed voxels along every ray match the original image. However, this is an optimization problem with immense complexity, analogous to stereo matching. Inspired by NeRF vs. stereo, we transform this discrete combination problem into a continuous optimization problem. Specifically, we select a color space for 3D printing pigments that supports linear superposition and includes density as a parameter. This allows pigments to mix linearly, similar to how RGB and density values are combined in VRPs to accumulate color along a ray. By integrating the Kubelka-Munk (K-M) model\cite{kubelka1931beitrag, Kubelka:48} with a pigment calibration process, we can predict the physical properties of pigment mixtures—such as their color and transmission rates—based on their absorption and scattering characteristics. This linear model enables us to simulate how mixed pigments behave in a local space, analogous to the way VRPs compute voxel contributions in the integration process along light paths.

In practice, we use a spectrophotometer to measure the reflectance and transmission rates of the six pigments across visible wavelengths, allowing us to determine their absorption and scattering properties through numerical solutions of the K-M model. Then, using this model, we compute the reflectance and absorption rates of mixed pigments, optimizing the mapping from RGB and density to ideal pigment concentrations. Since inverse mapping lacks an analytical solution, we extend methods like Mixbox\cite{sochorova2021practical} to establish a printable color space that allows rapid querying of pigment concentrations for any target color. Finally, recognizing that the printable pigment concentrations are discrete rather than continuous as calculated, we design a 3D stochastic halftoning method, inspired by 2D halftoning, to sample pigments based on their ideal concentrations. This method produces the final pigment label for each voxel, enabling precise and high-fidelity printing of volumetric models.

Our results demonstrate that DreamPrinting can accurately reproduce the visual effects of radiance representations with exceptional color fidelity and precision. VPPs facilitate the creation of translucent or semi-transparent objects with consistent internal properties, excelling in the reproduction of fine details such as leaves, fur, and clouds. By leveraging volumetric data, VPPs manage translucency more effectively than surface-based methods, enabling the accurate reproduction of transparent materials and intricate internal structures. We further validate the effectiveness of our pigment modeling and optimization process through comparisons with baseline approaches, highlighting significant improvements in both visual quality and computational efficiency. Additionally, we showcase the seamless integration of VPPs with emerging 3D generation techniques, such as \textit{TRELLIS}~\cite{xiang2024structured}. This integration underscores the versatility of VPPs, paving the way for more advanced and aesthetically refined printed objects.
\section{Related Work}

\subsection{Radiance Fields}

The volume rendering technology based on Radiance Field has demonstrated remarkable visual effects in film and gaming, capable of realistically simulating complex natural phenomena such as flames and clouds\cite{stam2023stable, dobashi2000simple, villemin2013practical, museth2013vdb, losasso2004simulating}. Traditionally, constructing these realistic Radiance Fields required substantial manual efforts from artists\cite{hasegawa2010love, murphy2018efficient}. Consequently, volume rendering was typically limited to these specialized scenarios.

With the advancements in neural network technology, volumetric data has demonstrated significant advantages in novel view synthesis. These advantages include the ability to preserve high-frequency details, flexibly represent complex geometric structures, and facilitate gradient-based optimization \cite{sitzmann2019deepvoxels, Lombardi:2019}. Neural Radiance Fields (NeRF) \cite{mildenhall2021nerf} pioneered the use of Multi-Layer Perceptrons (MLPs) as volumetric rendering primitives (VRPs), substantially enhancing the quality of novel view synthesis. Following this breakthrough, numerous studies have sought to optimize the structure of VRPs to reduce computational costs and improve rendering efficiency \cite{fridovich2022plenoxels, mueller2022instant, chen2022tensorf}. Additionally, recent research has extended the capabilities of VRPs to encompass material and lighting properties of real-world objects, enabling more realistic and physically accurate representations \cite{verbin2022refnerf, liang2023envidr, jiang2023gaussianshader}.

Concurrently, 3D generation\cite{gao2022get3d} integrates NeRF into generative adversarial networks (GANs) to achieve 3D-aware image synthesis\cite{chan2022efficient, chan2021pi, schwarz2020graf, niemeyer2021giraffe, gu2021stylenerf}. Diffusion models\cite{ho2020denoising, sohl2015deep} enhanced the generation quality and extended to voxel grids \cite{hui2022neural, muller2023diffrf, tang2023volumediffusion}. Notably, \cite{xiang2024structured} has introduced a text-to-3D generation method based on transformer, which by training on extensive and diverse 3D asset datasets, facilitates high-quality 3D asset generation and supports the output of Radiance Fields.

\subsection{Full-Color 3D Printing}
Among the various solutions for full-color 3D printing, Voxel Droplet Jetting technology stands out for its ability to reproduce highly detailed colors and intricate structures. One of the most representative implementations of this technology is the Stratasys PolyJet series of printers~\cite{stratasysJ850Prime}. These printers achieve complex color gradients and rich textural effects by precisely dispensing different pigment droplets at specific spatial locations\cite{Yuan2021AccurateAC}.

Currently, 3D printing technologies primarily focus on surface color printing, addressing challenges in color reproduction, scattering, and surface optimization. 
Error diffusion halftoning methods \cite{Brunton_Arikan_Urban_2015} have been effective in reducing color distortion in translucent materials, enhancing the fidelity of texture mesh printing.
To further improve clarity, \cite{Elek2017ScatteringawareTR} proposed scattering-aware techniques optimizing material distribution to minimize light scattering.
Structural-aware halftoning \cite{Abedini2022StructureAwareHU, Abedini20232DA3} focused on preserving sharpness and surface structural features. 
\cite{Babaei_Vidimče_Foshey_Kaspar_Didyk_Matusik_2017} optimized ink concentration to prevent spatial artifacts in continuous tone printing. Additionally, advances in material diffusion modeling enhance surface appearance predictions \cite{Luci2024DifferentiableMO}, and computational fabrication enables high-quality, personalized designs like color tattoos on varied skin tones \cite{Piovari2023SkinScreenAC}.

\cite{Nindel2021AGF} introduced the concept of volume rendering to 3D printing for the first time, proposing the use of volume parameterization to enhance the richness of surface color expression. However, this approach relies on an "opaque color" for the interior structure, making it challenging to directly apply to the fabrication of objects with translucent structures, such as those based on radiance representation.

\subsection{Color Science and the K-M Model}

Real-world inks reflect and scatter light when illuminated, ultimately producing a translucent effect with color \cite{franz2016colour}. To describe this behavior, we leverage the Kubelka-Munk (K-M) model \cite{kubelka1931beitrag, Kubelka:48}, which characterizes materials based on their absorption and scattering coefficients. These properties allow for the calculation of an object's response to light across different wavelengths. Early work by \cite{abdul2005spectral} first applied the K-M model to establish volume rendering for uniform materials, enabling a more accurate approximation of real-world color changes.

Duncan \cite{duncan1940colour} further explored the K-M model, demonstrating that the absorption (K) and scattering (S) coefficients change linearly when pigments are mixed. Based on these principles, Mixbox \cite{sochorova2021practical} implemented a method to convert digital colors into real-world pigment mixing ratios. Additionally, research in 3D printing and manufacturing \cite{piovarci2023skin, lee2024theory, Nindel2021AGF, Babaei_Vidimče_Foshey_Kaspar_Didyk_Matusik_2017, ansari2020mixed} has investigated the effects of material mixing on color properties, aiming to develop improved color reproduction techniques. \cite{papas2013fabricating, sitthi2015multifab, brunton20183d, shi2018deep} further improved color and transparency accuracy on translucent objects.

\section{Volumetric Printing Primitives}
\subsection{From Volumetric Rendering to Volumetric Printing}
\label{sec:vpp}

\begin{figure}
  \includegraphics[width=\linewidth]{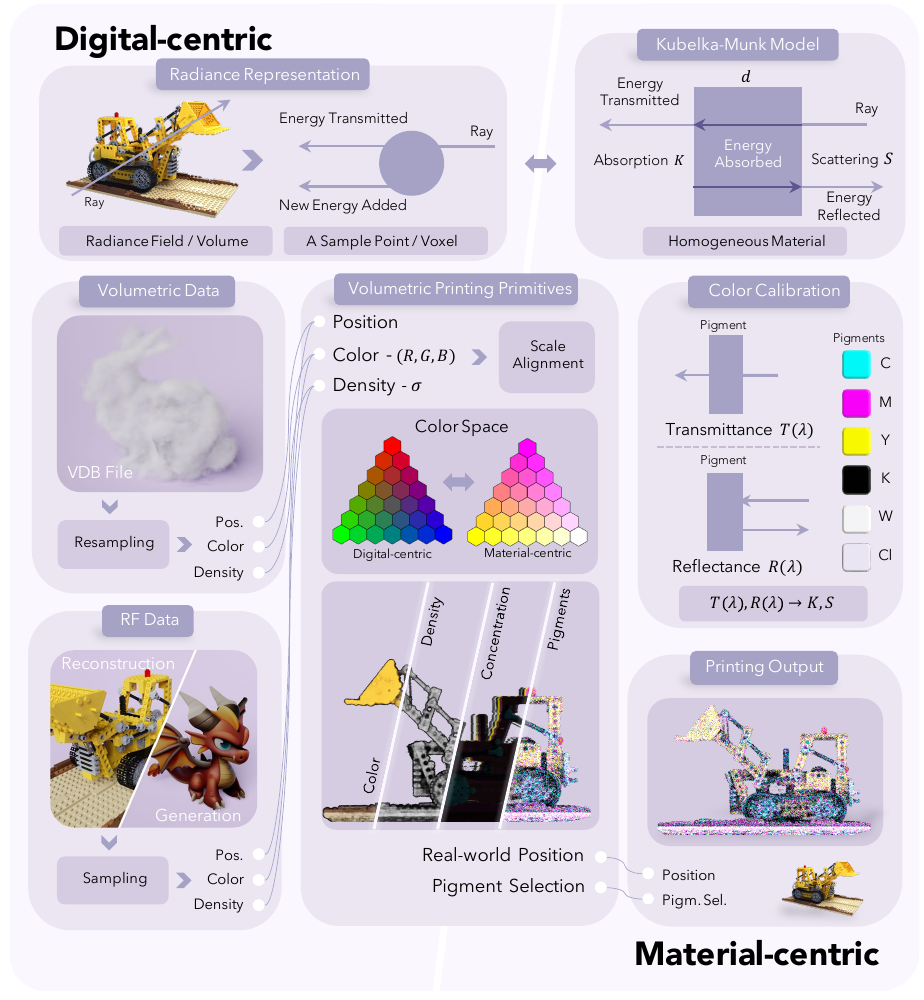}
  \caption{We employ the Volumetric Printing Primitive to bridge the gap between radiance representation and material-centric representation. Using the Kubelka-Munk model, we simulate the pigment mixing process, which serves as the foundation for color calibration and the construction of a printing color gamut with efficient lookup capabilities. This enables the rapid conversion from RGB and density values to pigment concentrations. Finally, we adopt halftoning to generate a print-ready pigment selection.}
  \label{fig:methods}
\vspace{-0.8cm}
\end{figure}

A volumetric printing primitive (VPP) converts a radiance representation—such as a radiance field or radiance volume—into a 3D printable, material-centric volume, as shown in Fig.~\ref{fig:methods}. Similar to volumetric rendering primitives (VRP), VPPs are used to describe spatial color and density distributions. However, unlike VRPs, where color and density values can be continuously and freely combined, the domain of VPPs is a discrete space. This limitation arises from the constraints of the 3D printer's nozzle architecture, which allows only a finite number of pigments to be used in each printing task. Furthermore, at any given spatial location, only a single pigment can be deposited.

In our experiments, we utilized one of the most advanced 3D printers, the Stratasys J850 Prime~\cite{stratasysJ850Prime}, which supports six primary pigments: Cyan ($\texttt{C}$), Magenta ($\texttt{M}$), Yellow ($\texttt{Y}$), Black (Key, $\texttt{K}$), White ($\texttt{W}$), and Clear ($\texttt{Cl}$).

To enable 3D printing, the first step is to sample the radiance representation. 
Recall that a radiance field is generally represented as a mapping function:
\begin{equation}
\mathcal{F}: \mathbb{R}^3 \to \mathbb{R}^4, \quad \mathcal{F}(x, y, z) = (r, g, b, \sigma),
\end{equation}
\noindent where \((x, y, z)\) denotes the spatial coordinates and \((r, g, b, \sigma)\) corresponds to the color and density values at the queried location and direction.

By sampling the radiance representation at the grid points corresponding to each volumetric printing primitive (VPP) $\mathbf{v}$, we obtain the RGB color values $\mathbf{r}^\dagger(\mathbf{v})$ and the density $\sigma^\dagger(\mathbf{v})$.

A brute-force approach considers a local spatial region, for example, treating a $3 \times 3 \times 3$ neighborhood around each voxel as a hyper-voxel. In the radiance representation, the average color of the region is computed by density-weighting the colors, and then approximated using a set of discrete pigment labels. The result of this process is illustrated in Fig.~\ref{fig:ablation_study}. However, the printed result may exhibit a dull appearance with excessive transparency and inaccurate color reproduction. This issue arises due to the lack of proper pigment modeling and inadequate calibration of pigment transparency and color properties. Furthermore, this approach incorrectly models the light interactions inherent to volume rendering, leading to discrepancies between the expected and actual printed appearance.

To accurately achieve the conversion from volumetric rendering primitives (VRP) to volumetric printing primitives (VPP), we designed a theoretical framework. First, we assume that each voxel can contain multiple pigments mixed at arbitrary concentrations:
\begin{equation}
    C = \{c_i \mid i \in I = \{\texttt{C,M,Y,K,W,Cl}\}, \sum_{i \in I} c_i=1, c_i\in[0,1]\},
\end{equation}
\noindent where $c_i$ indicates the concentration of the pigment $i$. This assumption allows us to transform the discrete optimization problem into a continuous one, which is easier to solve. Next, we build upon the Kubelka-Munk (K-M) model to establish a mapping from pigment concentrations to RGB color and density values. This mapping method, centered on the absorption ($K$) and scattering ($S$) coefficients of the pigments, accurately simulates the physical properties of pigment mixtures within a small local region.

To achieve the inverse mapping from RGB color and density to the K-M model parameters, we solve the inverse K-M problem. However, since this inverse process involves transcendental equations, obtaining an analytical solution is infeasible. To address this, we extend the method proposed in~\cite{sochorova2021practical} for rapidly mapping opaque pigment concentrations to RGB values. By calibrating the pigments, we establish a printable color space that allows for rapid lookup of the required pigment concentration combinations to achieve a target color.

Once the optimal pigment concentrations are determined, we adopt a 3D stochastic halftoning technique inspired by 2D halftoning. This method performs random sampling based on the ideal pigment concentrations, ultimately yielding the final pigment label.

\subsection{ Physical Modeling with the Kubelka-Munk Framework}

We adapt the Kubelka-Munk (K-M) model to characterize the physical properties of pigments. Given a concentration $C$, the reflectance $R_C(\lambda)$ and transmittance $T_C(\lambda)$ can be computed using the K-M model, which expresses them as functions of the absorption $K_C(\lambda)$ and the scattering $S_C(\lambda)$ at a specific wavelength $\lambda$. The reflectance and transmittance at a given thickness $t$ are defined as follows~\cite{abdul2005spectral}:

\begin{align}
\label{eq:R_lambda}
R_C(\lambda) &= \frac{\sinh(b S_C(\lambda) t)}{a \sinh(b S_C(\lambda) t) + b \cosh(b S_C(\lambda) t)},  
\\
T_C(\lambda) &= \frac{b}{a \sinh(b S_C(\lambda) t) + b \cosh(b S_C(\lambda) t)},
\label{eq:T_lambda}
\end{align}

\noindent where 
\[
a = \frac{S_C(\lambda) + K_C(\lambda)}{S_C(\lambda)}, \quad b = \sqrt{a^2 - 1}.
\]

Unlike the digital domain, where color mixing is typically represented as a weighted average of RGB values, pigment mixing in the real world follows a different principle. Instead of directly averaging colors, the absorption and scattering coefficients, denoted by $K$ and $S$, respectively, are combined using a weighted sum. The following equations describe how multiple pigments collectively contribute to absorption and scattering within a small region, enabling the computation of reflectance and transmittance:

\begin{equation}
K_{C}(\lambda) = \sum_{i \in I} c_i K_i(\lambda),  \quad S_{C}(\lambda) = \sum_{i \in I} c_i S_i(\lambda),
\end{equation}

\noindent where $c_i$ represents the concentration of pigment $i$ in the mixture, and $I$ denotes the set of available pigments.

Building upon the K-M model, we further propose an approximation for real-world materials using three-channel RGB values and a scalar density. This approach enables correspondence with the radiance representation used in rendering.
In this method, the observed color is decomposed into two components: reflected light ($R_C$) and transmitted light ($T_C$). We use the CIE 1931 color model under the assumption of uniform white light illumination ($D_{65}$) as our color mapping function ($\mathbf{CMF}$). The RGB color can be computed from the spectral power distribution as follows:
\begin{equation}
\label{eq:cmf}
    \mathbf{r}_C = \mathbf{CMF}(\{ (R_C(\lambda) + T_C(\lambda)) \mid \lambda \in \Lambda \}, D_{65})
\end{equation}
\noindent where $R_C(\lambda)$ and $T_C(\lambda)$ represent the reflectance and transmittance ratios at a specific wavelength $\lambda$, and $\Lambda$ denotes the human-visible light spectrum within the range $[380, 750]$ nm. The function $\mathbf{CMF}$ is a standard transformation that converts spectral intensity data into RGB values by weighting the contributions of different wavelengths based on the human visual system's sensitivity under the specified illuminant $D_{65}$.

For density estimation, we observe that although Eq.~\ref{eq:T_lambda} has a complex expression when the ratio of scattering coefficient $S$ to absorption coefficient $K$ is relatively small, the function can be well approximated by the commonly used volume rendering opacity function, $\exp (-\sigma \cdot t)$. Therefore, we determine the optimal value of $\hat{\sigma}_C(\lambda)$ that best fits the transmittance $T$ using the radiance representation for each wavelength $\lambda$ as follows:
\begin{equation}
\hat{\sigma}_C(\lambda) = 
\underset{\sigma}{\arg \min} \int_0^{t_{\text{max}}} \left\| \exp(-\sigma \cdot t) - T_C(\lambda, t) \right\|^2 \, dt
\label{eq:average_transmission}
\end{equation}
\noindent where $t_{\text{max}}$ is a manually defined, sufficiently large value that ensures $T_C(\lambda, t_{\text{max}})$ approaches zero, facilitating an accurate solution. The effectiveness of this approximation is demonstrated in Tab.~\ref{tab:pigment_errors}.

Next, we define the scalar $\sigma_C$ by averaging the transmittance across all measured wavelength bands:
\begin{equation}
\label{eq:sigma_c}
    \sigma_{C} = \ln \left( \frac{1}{\| \Lambda \|} \sum_{\lambda \in \Lambda} \exp \left( - \hat{\sigma}_C (\lambda \cdot \Delta t) \right) \right) \Big/ \Delta t
\end{equation}
\noindent where $\Lambda$ represents the set of all measured visible light bands, and $\| \Lambda \|$ denotes the total number of bands in the set.

\begin{figure}[ht]
    \centering
    \includegraphics[width=\linewidth]{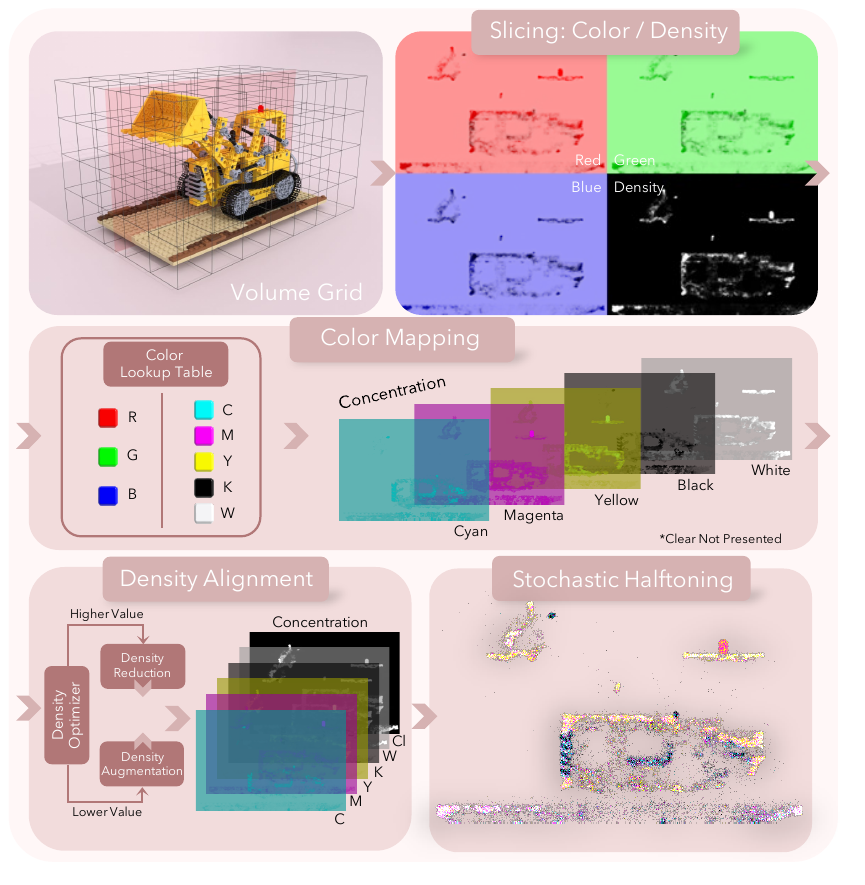}\
    \caption{Leveraging the properties of VPPs, all operations can be executed in parallel. For each voxel in the radiance representation, expressed in terms of RGB and density, pigment concentrations of \texttt{C}, \texttt{M}, \texttt{Y}, \texttt{K}, and \texttt{W} are determined using a pre-computed color lookup table. Subsequently, a density alignment strategy is applied to adjust the printing density, ensuring alignment with the radiance representation. Finally, stochastic halftoning is employed to generate the print-ready pigment distribution.
}
    \label{fig:slicing}
\vspace{-0.5cm}
\end{figure}

\section{Concentration Optimization}

\begin{figure*}
  \includegraphics[width=0.85\linewidth]{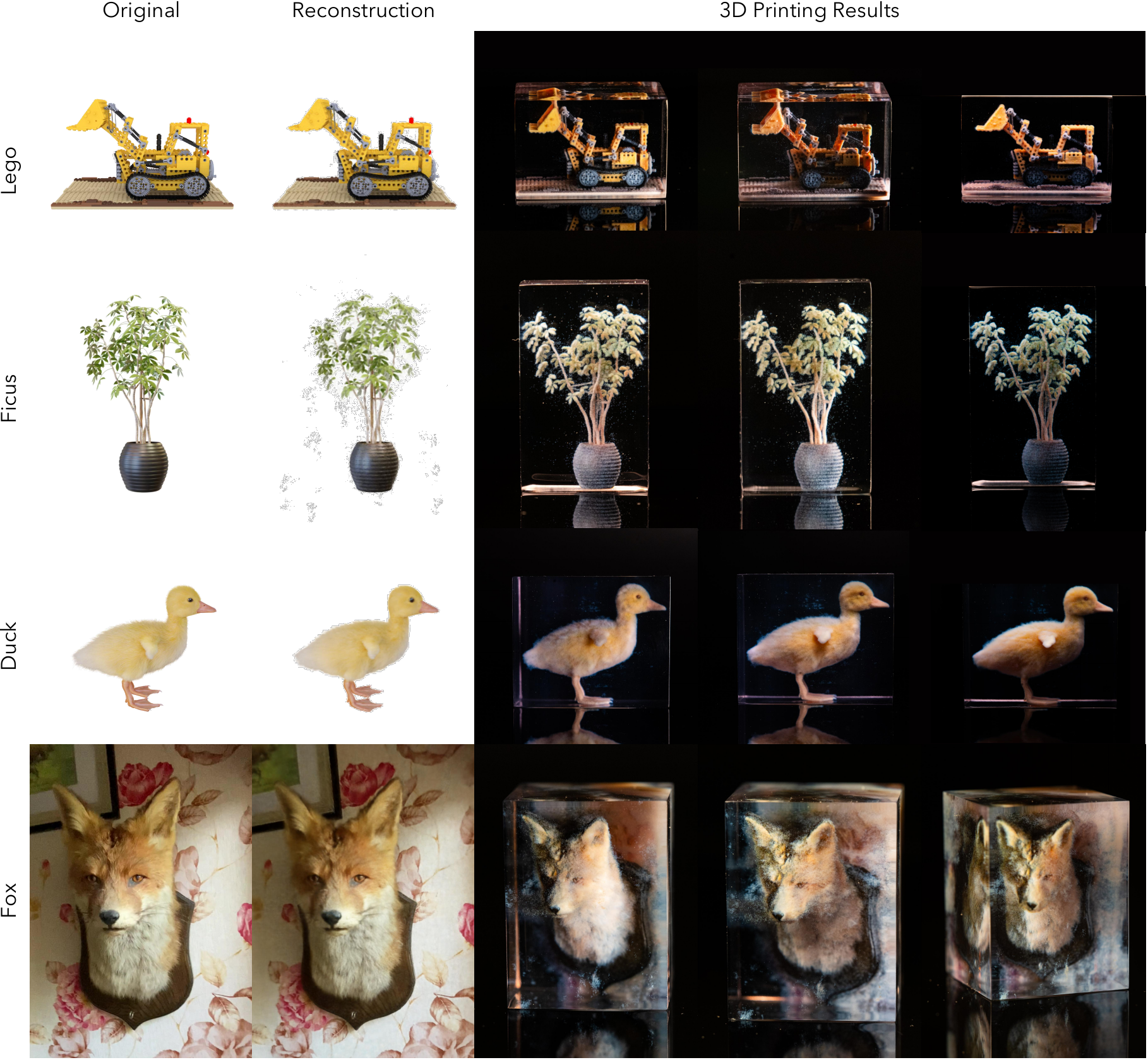}
\caption{
Our results gallery of 3D printing objects from InstantNGP\cite{mueller2022instant} radiance reconstructions. From left to right: the original images, the 2D images rendering from the radiance reconstruction, and the corresponding 3D printing results. 
}
  \label{fig:gallery}
\end{figure*}

\subsection{Concentration Optimization and Density Alignment}
\label{sec:sec4.1}

Our goal is to determine a pigment concentration combination $C$ such that the printed color $\mathbf{r}_C$ closely approximates the target RGB color $\mathbf{r}^\dagger$ and density $\sigma^\dagger$ obtained from the radiance representation. 

To achieve this, we choose not to simultaneously optimize both color and density due to two primary considerations. First, color and density exist on different scales, making it challenging to design a unified optimization objective. Second, the presence of transparent pigments introduces complexities, as direct simultaneous optimization of color and transparency can lead to an ill-posed problem. Given that the impact of transparent pigments on color is minimal, we propose a two-stage approach: initially, we exclude the Clear pigment \texttt{Cl} and optimize the concentrations of the remaining five pigments. Subsequently, we incorporate the Clear pigment to achieve density alignment.

In the first stage, the optimal concentration $C^*$ is obtained by solving the following optimization problem:

\begin{equation}
\label{eq:least_squares_color}
C^*(\mathbf{v}) = \underset{C}{\arg \min} \left\| \mathbf{r}_C - \mathbf{r}^\dagger(\mathbf{v})\right\|_2^2, 
\end{equation}

\noindent $\text{s.t.}$
\[
C = \{c_i \mid i \in I = \{\texttt{C,M,Y,K,W,Cl}\},  \sum_{i \in I} c_i=1,  c_i \in [0,1],  c_{\text{Cl}} = 0\},
\]

Using Eq.~\ref{eq:sigma_c}, we obtain the density $\sigma_{C^*}$ for the current pigment concentration combination $C^*$. 
To achieve density alignment with the target value $\sigma^\dagger$, we tune the concentration of Clear or Black/White pigment.

If the computed concentration $C^*$ results in $\sigma_{C^*} > \sigma^\dagger$, the overall density is reduced by incorporating a proportion $\rho^-$ of the Clear (\texttt{Cl}) pigment into $C^*$. Since the absorption coefficient $K_{\text{Cl}}$ and the scattering coefficient $S_{\text{Cl}}$ are approximately zero, the adjusted concentration $C$ is calculated as:
\begin{equation}
    C = \rho^-\cdot C^* + (1 - \rho^-)\cdot C_{\texttt{Cl}},
\end{equation}
\noindent where $\rho^- = \frac{\sigma^\dagger}{\sigma_{C^*}}$, and $C_{\texttt{Cl}}$ represents the concentration of the Clear pigment, with all other pigment concentrations set to zero, satisfying the condition $c_{\texttt{Cl}} = 1$.

For certain colored pigments, their high transmittance at specific wavelengths allows light to pass through even at high concentrations. To achieve a higher density for the resulting mixed color, it is necessary to introduce pigments that provide effective attenuation across all wavelength bands. In our case, the high-density pigments used for this purpose are Black \texttt{K} and White \texttt{W}.

Therefore, in the condition that concentration $C^*$ results in $\sigma_{C^*} < \sigma^{\dagger}$, we first combine \texttt{K} and \texttt{W}, which intuitively produce a gray color, and define this mixture as a new virtual pigment \texttt{X}. We then incorporate a proportion $\rho^+$ of \texttt{X} into $C^*$ to increase the overall density. However, it is important to note that introducing additional non-transparent pigments introduces a trade-off between achieving the desired density and maintaining color accuracy.

To determine the optimal mixing ratio $\rho^{\texttt{K}}$ of the \texttt{K} and \texttt{W} pigments, we solve the following optimization problem to ensure that the brightness of the mixture matches the brightness of the target RGB color $\mathbf{r}^\dagger$:
\begin{equation}
\label{eq:rhok}
\rho^{\texttt{K}} = \arg \min_\rho \left\| \overline{\mathbf{r}_{C_{\texttt{X}}(\rho)}} - \overline{\mathbf{r}^\dagger} \right\|_2^2,
\end{equation}
\noindent where $\overline{r}$ denotes the mean value of the RGB channels, representing brightness. The concentration $C_{\texttt{X}}(\rho)$ is defined as:
\[
c_i \in C_{\texttt{X}}(\rho) =
\{ \rho \, | \, i = \texttt{K}, \quad 1 - \rho \, | \, i = \texttt{W}, \quad 0 \, | \, \text{otherwise} \}
\]
The final adjusted concentration $C$ is then computed:
\begin{equation}
    C = \rho^+ \cdot C^* + (1 - \rho^+)\cdot C_{\texttt{X}}.
\end{equation}
where $C_{\texttt{X}}$ is a simplified notation for $C_{\texttt{X}}(\rho^\texttt{K})$, and in practice, to ensure color accuracy, we set the upper limit of $\rho^+$ to $0.1$.

\subsection{Constructing Efficient Printing Color Gamut}

A crucial aspect of applying the Kubelka-Munk (K-M) model is the requirement for each pigment $i$ to have corresponding absorption ($K$) and scattering ($S$) coefficients defined across all wavelengths in the visible spectrum $\lambda \in \Lambda$. We use a spectrophotometer to measure the reflectance $R_i(\lambda)$ and transmittance $T_i(\lambda)$ of each of the six pigments $i$ across the visible light bands $\lambda \in \Lambda$. By solving the inverse problem of Equations ~\ref{eq:R_lambda} and ~\ref{eq:T_lambda}, we can calculate the absorption coefficient ($K_i(\lambda)$) and scattering coefficient ($S_i(\lambda)$) of each pigment for each wavelength. We employ the Levenberg-Marquardt (LM) algorithm~\cite{more2006levenberg} for this inversion process.

With the calibration data, we can determine the corresponding RGB value and density for any pigment concentration combination $C$ using Eq.~\ref{eq:cmf} and Eq.~\ref{eq:sigma_c}. However, solving Eq.~\ref{eq:rhok} and Eq.~\ref{eq:least_squares_color} requires inverting the K-M model, which involves transcendental equations that lack analytical solutions. This computation is costly, especially for volumetric data structures with a large number of voxels.

To address this, we construct lookup tables for RGB-to-concentration and brightness-to-$\rho^\texttt{K}$ mappings. The table densities are $100^3$ and $100$, respectively. Using these tables, we perform linear interpolation to obtain the required concentration mapping for any RGB value. Due to the structure of VPPs, this process can be executed in parallel across each sliced layer.

\subsection{3D Stochastic Halftoning }

We adapt halftoning strategies from 2D printing to develop a method for 3D volumes. We call it 3D stochastic halftoning. Given the target color concentration $C$ for each VPP, we use it as the sampling probability to sample the label, ultimately obtaining the ink $\text{pigment}(\mathbf{v})$ to be printed on this VPP $\mathbf{v}$.
The halftoning results are print-ready, which is illustrated in Fig.~\ref{fig:slicing}. Each voxel is assigned a unique pigment label.

\section{Experiments}

\begin{figure}
    \centering
    \includegraphics[width=\linewidth]{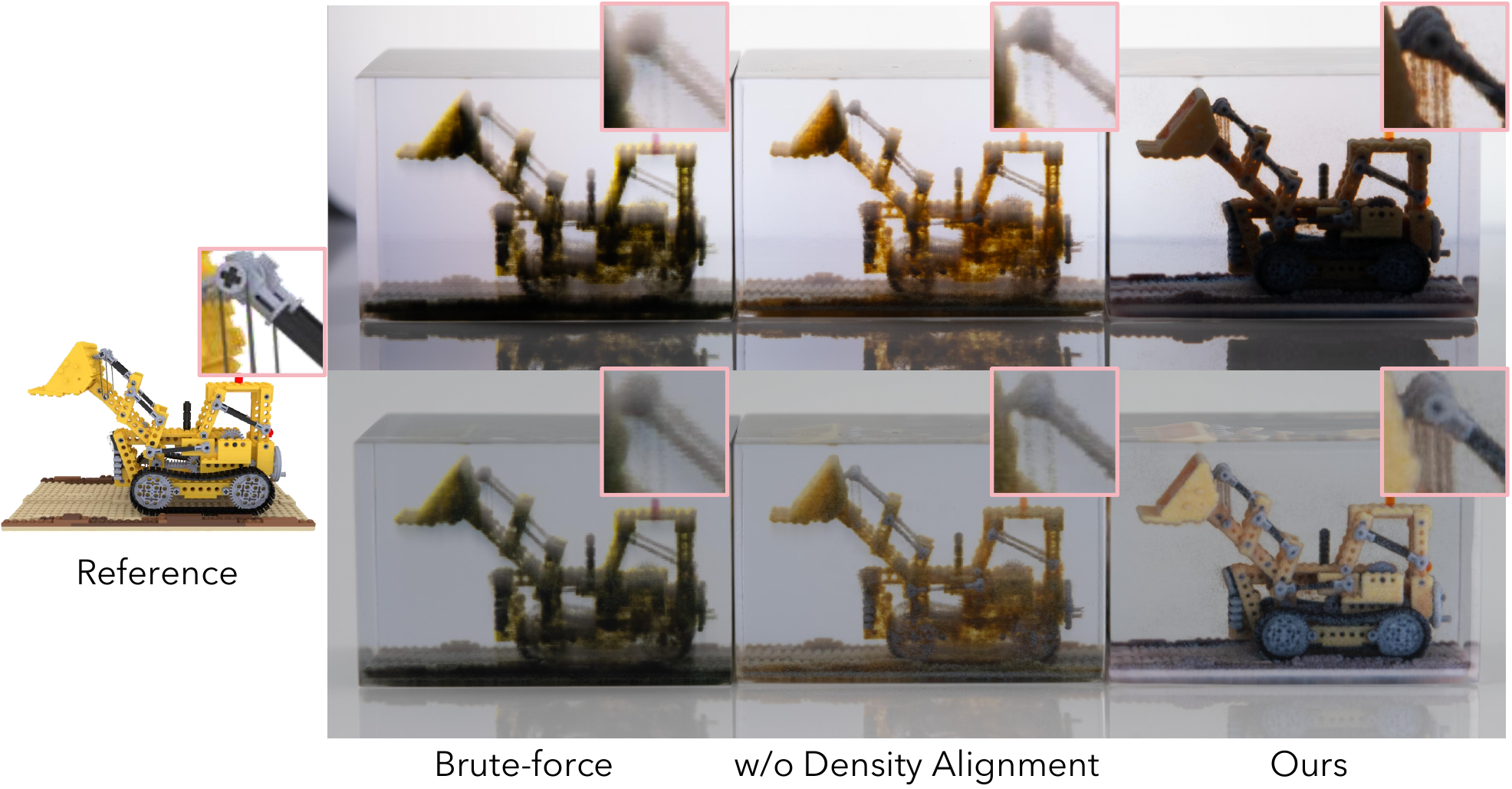}
    \caption{Ablation study. ``Brute-force'' and ``w/o Density Alignment'' suffer from incomplete opacity. Our method produces accurate opacity, which demonstrates the effectiveness of our density alignment strategy.}
    \label{fig:ablation_study}
\end{figure}

\begin{figure}
    \centering
    \includegraphics[width=\linewidth]{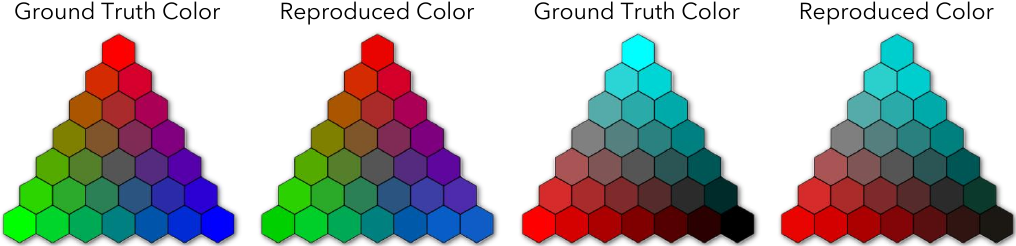}
    \caption{Our method accurately reproduces the target colors within the gamut of the pigments.}
    \label{fig:color_reproducing}
\end{figure}

\subsection{Implementation}
In this paper, we use the Stratasys J850 Prime 3D Printer\cite{stratasysJ850Prime} and Vero Vivid series materials\cite{VeroM} from stratasys. The physical size of a voxel is $[0.084, 0.028, 0.014]\text{mm}$. 
All experiments are conducted on Ubuntu 22.04, equipped with an Intel(R) Xeon(R) W-2223 processor and an NVIDIA GeForce RTX 3090 graphics card.
We use the Agilent Technologies Cary Series UV-Vis-NIR Spectrophotometer (Cary 5000)\cite{AgilentTechnologies} to perform molecular analysis. This instrument operates in a wavelength range of 175-3300 nm and includes a PbSmart near-infrared detector.

\subsection{3D Printing Object}
\begin{figure*}
  \includegraphics[width=0.9\linewidth]{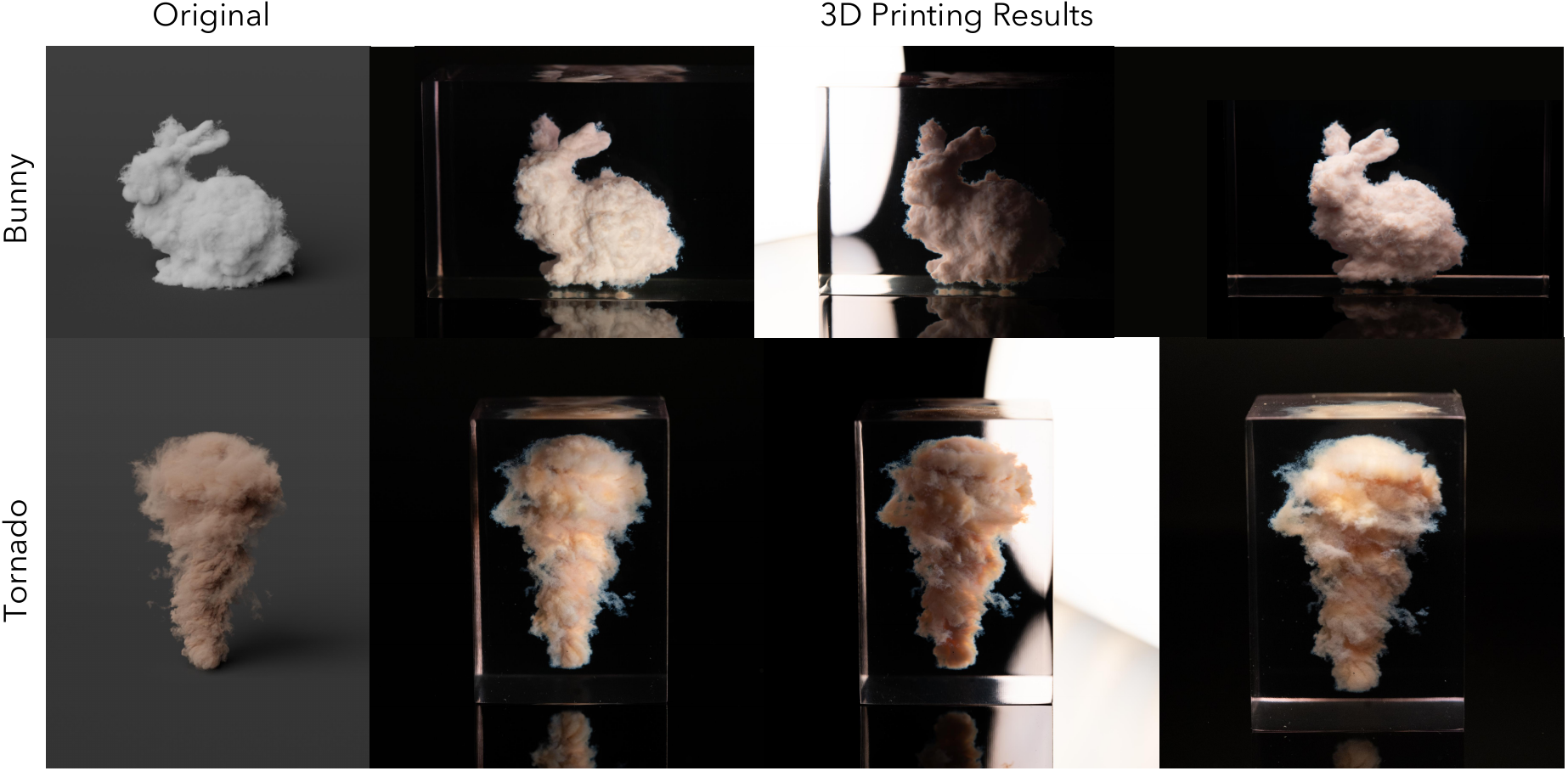}
  \caption{
  Gallery of artist-made OpenVDB assets. From left to right: the original volumetric representation, and the corresponding 3D printing results. Our method accurately replicates fine volumetric details and translucency effects.
  }
  \label{fig:gallery2}
\end{figure*}

We showcase a diverse collection of 3D printing results derived from three distinct radiance sources: artist-made, multi-view image reconstruction to image-to-3D model generation. 
As shown in Fig.~\ref{fig:gallery2}, we present the artist-made volumes. We extract per-voxel density from OpenVDB data and assign a uniform color to all voxels. The \textsc{Tornado}\cite{jangafx2020} cloud conveys fluffy textures, and \textsc{Tornado}\cite{jangafx2020} emphasizes a sense of motion. 

\begin{figure*}
  \includegraphics[width=0.9\linewidth]{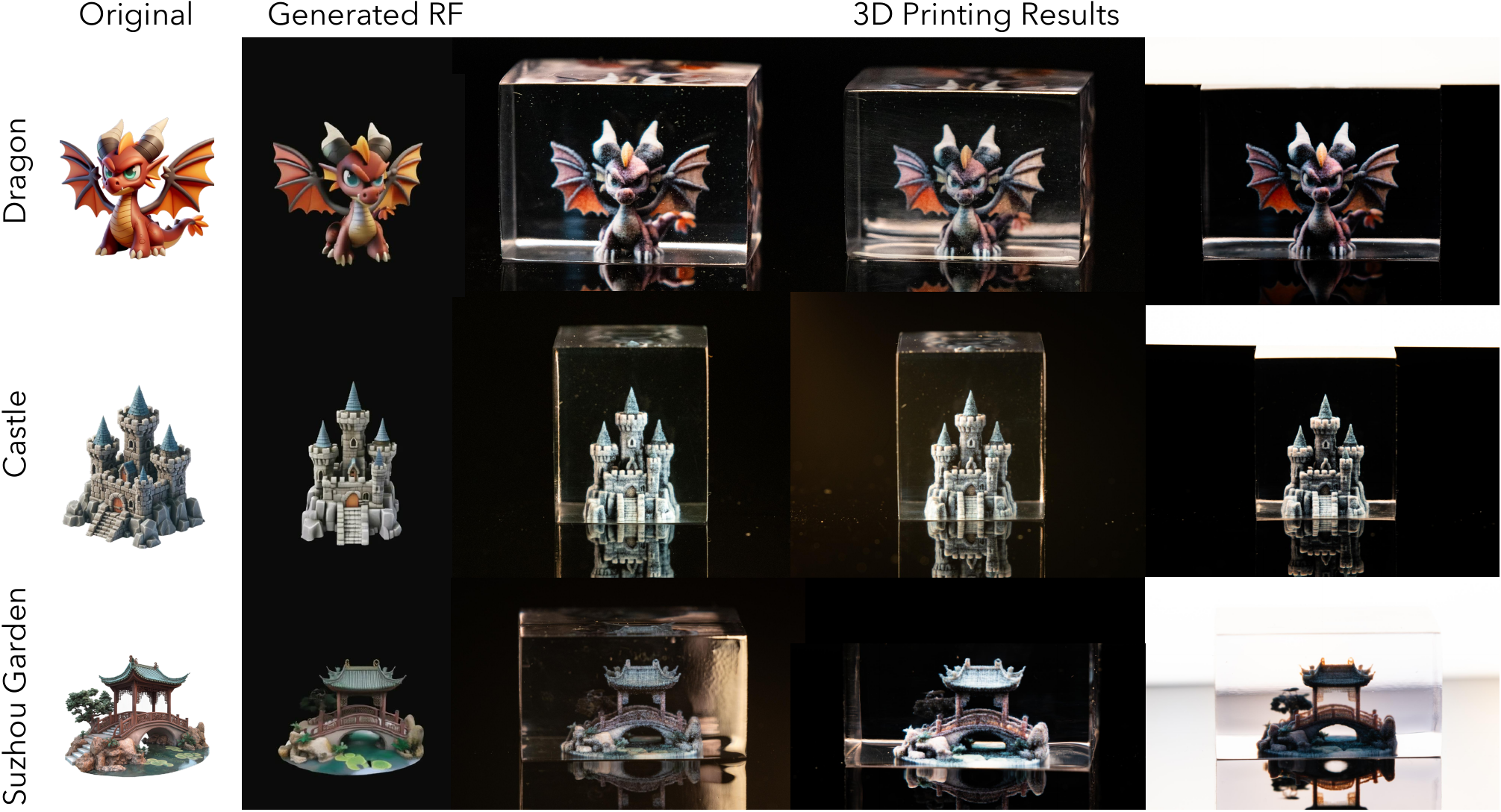}
  \caption{
Printings of radiance models generated by TRELLIS\cite{xiang2024structured}. From left to right: the prompt image, the generated radiance fields (RF), and the corresponding 3D-printed results. Our approach faithfully captures intricate color details and spatial structures, demonstrating high-fidelity reproduction in physical form.}
  \label{fig:gallery3}
\end{figure*}

For multi-view image reconstruction as shown in Fig.~\ref{fig:gallery}, we use InstantNGP\cite{mueller2022instant} to reconstruct the radiance field from multi-view image inputs.
We present the printed results from the well-known image dataset NeRF Synthetic\cite{mildenhall2021nerf}, such as \textsc{lego} and \textsc{ficus}, which feature complex geometries. Our approach faithfully captures fine details like hinges, connecting rods, and blades, enabling the creation of highly accurate small-scale objects with exceptional precision in both geometry and color.
In addition to these solid objects, we printed furry animals reconstructed from image data. \textsc{duck}\cite{luo2022artemis} vividly showcases intricate details with varying volume densities, effectively simulating textures like soft feathers and hard beaks. For the \textsc{fox}~\cite{mueller2022instant} dataset, the training images cover a limited range of views, unlike other cases where 360-degree views around the object are available. Our method successfully reproduces the same visual effects as the radiance field, even in the presence of floaters and suboptimal reconstruction results.

For image-to-3D model generation, we present results produced by TRELLIS\cite{xiang2024structured} as illustrated in Fig.~\ref{fig:gallery3}, including mythical characters \textsc{Dragon}, \textsc{Castle}, and \textsc{Suzhou garden}. Our method bridges the gap between imagination and reality.
We present prints under multiple perspectives and lighting conditions, revealing how different illuminations accentuate the materials of each print.

\subsection{Evaluation}

We evaluated the accuracy of our color and density modeling approach. Subsequently, we assessed the necessity of the Density Alignment strategy.
\vspace{-0.2cm}

\paragraph{Color Reproducing}
In Fig.~\ref{fig:color_reproducing}, we compare the printing color gamut with the sRGB color gamut. 
For each hexagonal grid, we use the lookup table to determine the pigment concentrations based on its RGB value, and reproduce the RGB value using the Kubelka-Munk (K-M) model. Our results show that, for the vast majority of colors, the pigments can accurately reproduce the target colors.

\begin{table}[h!]
\centering
\caption{Error of approximating $T(t)$ with $\exp(-\sigma \cdot t)$.  The ``Mean'' row represents the average error of this approximation over all thickness $t$ samplings across all wavelengths. The ``Max'' row represents the average error over all thickness samplings under the worst wavelength.}
\vspace{-0.2cm}
\begin{tabular}{l|cccccc}
\hline
\textbf{Pigment} & \texttt{C} & \texttt{M} & \texttt{Y} & \texttt{K} & \texttt{W} \\ \hline
Mean & 
0.017\% & 0.026\% & 0.020\% & 0.000\% & 0.847\% \\ 
Max & 
0.251\% & 0.082\% & 0.048\% & 0.000\% & 1.898\% \\ \hline
\end{tabular}

\label{tab:pigment_errors}
\vspace{-0.6cm}
\end{table}

\paragraph{Density Approximation}
We quantitatively verify the validity of the pigment density approximation method described in Eq.~\ref{eq:average_transmission}.
For a given wavelength $\lambda$, we uniformly sample 100 thickness values $t$ within the range $[0, t_{\text{max}}]$ and compute the transmittance $T(\lambda,t)$ using Eq.~\ref{eq:T_lambda}, where $t_{max}$ is set to $5 \text{mm}$ in this experiment. The error is calculated as the difference between the fitted value $\exp(-\sigma \cdot t)$ and the actual value $T(\lambda,t)$, and then averaged over the sampled thickness values. 
This evaluation process is performed across the visible wavelength range for each pigment. To assess the approximation quality, we report two metrics: the average and the maximum error within all wavelengths.
As shown in Tab.~\ref{tab:pigment_errors}, the errors are minimal, confirming the validity of our proposed density approximation method on our selected pigments.

\paragraph{Density Augmentation Strategy}
We further validate the density augmentation strategy described in  Sec.~\ref{sec:sec4.1}.
As illustrated in Fig.~\ref{fig:ablation_study}, "w/o density alignment" refers to the absence of this strategy. Instead, both RGB and density are used to solve the optimization problem. This approach is more time-consuming, and because it is difficult to adjust the weights between density and RGB, the final density result does not closely approximate the theoretical value. ``brute-force'' refers to the results described in Sec.\ref{sec:vpp}. We observed them under different lighting conditions. 
This underscores the critical role of density-adjustment strategies in achieving an optimal balance between opacity and color fidelity in 3D-printed models.

\section{Conclusion}

In this paper, we propose the Volumetric Printing Primitive (VPP), which bridges the gap between volumetric rendering and volumetric printing. Our approach transforms discrete printing pigment labels into continuous representations and models translucent pigment mixing based on the Kubelka-Munk (K-M) theory to accurately simulate the color and density of mixed pigments in local spatial regions. In practice, we construct color spaces for translucent pigments that enable the rapid mapping of RGB and density values to pigment concentrations, and we generate printable pigment distributions using a stochastic halftoning method.

The results demonstrate that our method can faithfully reproduce visual effects captured in radiance representations, such as ficus leaves, furry animals, and clouds, with a high degree of accuracy. Furthermore, we showcase the potential of our approach to achieve highly detailed and realistic results when integrated with emerging 3D content generation techniques.

Our work opens up new possibilities for enhancing the aesthetics and realism of 3D-printed objects by enabling more accurate reproduction of color and translucency. This advancement has potential applications in fields such as art, design, visualization, and scientific modeling, where precise appearance reproduction is critical.

\bibliographystyle{ACM-Reference-Format}
\bibliography{sample-bibliography}

\end{document}